# Spectra of quark-antiquark bound states via two derived QCD potential

**M. S. Ali\*, A. M. Yasser**
African Institute for Mathematical Sciences, B.P. 1418 Mbour, Thiès Region, Senegal.
Physics Department, Faculty of Science at Qena, South Valley University, Egypt.

*E-mail : <mohamed.s.g.ali@aims-senegal.org>

E-mail : <yasser.mostafa@sci.svu.edu.eg>

**Abstract** : In the current paper, we propose two types of quark-antiquark $Q\bar{Q}$ interactions, which may be tailored to describe various meson sectors. The interactions contain Quantum Chromodynamics (QCD) inspired components, such as the Coulomb-like interaction, the confinement linear potential, and the spin-spin interaction. Our scheme relies on the non-relativistic quark model through the introduction of two derived QCD potential models. The application of the two proposed potentials resulted in spectra for $Q\bar{Q}$ bound states, which are compared with published experimental data. We found that one of the two potentials is favored over the other in terms of high precision comparisons.

## 1. Introduction

Since the birth of quark model in 1964, meson properties have been extensively studied. Meson is a quark-antiquark configuration. The main motivation from investigating meson properties in quark model is to understand the model applicability and generate possible improvements. There are many studies on the meson transitions, for example but not limited those studies in Kwong et al [1, 2], Godfery and Rosner [3, 4], J. Segovia et al [5, 6], and Kumar et al [7].

Certain modifications to the model are suggested which have been inspired by fundamental Quantum Chromodynamics essential properties, such as spin dependence of strong interactions.

There are various models proposed to describe the hadronic sector. Through these models, the $Q\bar{Q}$ potential models have proven to be successful. One of the most reliable models is the models which developed by Godfrey and Isgur (GI) in 1985 [8]. Its description of both spectra and decay properties of mesons relies on relativized properties of the kinetic and potential terms. Appreciably, the energy terms in this model have included most of basic ingredients of QCD.

Following this approach, many attempts try to step towards more sophisticated treatments of GI model through sticking more to the basic properties of QCD [5-6]. The most dependable development is the work done by C. Semay and B. Silvester-Brac (CB) in 1997 [9].

Actually, the purpose of our work is manifold. We want to study the spectra of heavy quark-antiquark bound states and illustrate their wavefunctions through enhancing the previous mentioned approach. We will do that via considering two phenomenological heavy quark-antiquark potentials that depends on requisite QCD features similar to those used by both GI and CB models, but avoiding its too considerable complexity.

Our new perspective relies on numerical treatment. Numerical treatment of our considered approach is much easier and faster. Consequently, it can extend to be applicable to more complex system without serious difficulties, such as baryons.
Indeed, a great variety of numerical techniques have been developed to study the hadronic sectors in both relativistic and non-relativistic quark model. In preliminary attempts, Lattice calculations played an important role. They are a reliable techniques for simple systems, but this numerical approach is extremely computationally expensive [10, 11]. Lattice calculations are lattice formulations of gauge theory which proposed in 1974. Until the present time, they still need great improvements to be applicable for complex system.

Besides, there are an additional techniques have been developed in both relativistic and non-relativistic frameworks [12, 13], such as, the Fourier grid Hamiltonian method [14], Shooting method [15], and Numerov method [16]. One of our aims from this work is using a numerical technique that achieves a good accuracy with avoiding the tremendous computer time which associated with the previous techniques of the preceding studies. In our perspective, Matrix method is considered.

From previous studies, although many forces are definitely effective in complex quark-antiquark systems, but the corresponding expression are not available yet [17]. Thus, simplicity dictates that we should consider only sum of two-body interactions. It worth noting that, a sum of two-body interaction is a good approximation for three-body interactions [18].

Genuinely, this approximation becomes much poorer for multi-quark systems as the number of quarks increases. We hope this approximation that gives the gross features of heavy quark-antiquark investigation. Moreover, R. Vinh Mau, C. Semay, et al have proven that many-body forces play a minor role for characterizing the hadron-hadron interactions [19].

Our philosophy is based on using phenomenological energy terms that is compatible with basic QCD ingredients. We use this phenomenological approach for two reasons. Firstly, we are interested in an overall description of heavy mesons spectra. Secondly, it is certainly able to treat our physical problem in much cheaper and satisfactory way than the preceding approaches.

Our work should be considered as a step providing a completely phenomenological ansatz, difficult to justify theoretically, which allows others in near future to treat complex systems with a good accuracy and with very short computer time.
As pointed out previously, we are interested in the gross features of heavy $Q\bar{Q}$ spectra. From this perspective, we shall restrict our attention on two potentials of quark-antiquark bound states while respecting a great deal of the fundamental Chromodynamics theory. Consequently, we ignore the spin-orbit and tensor forces to gain some simplicity.

Indeed, we are aware that this simplification is for a realistic potential, but this work must be considered as a preliminary necessary step before a more sophisticated $Q\bar{Q}$ description can be given including these corrections.

Formerly, some attempts make a very precise comparison between the use of non-relativistic and relativistic frameworks have been done.

Our paper is organized as follows, in next section we present the model Hamiltonian associated with two phenomenological derived QCD potentials. In the third section, we explain the numerical scheme that used to study $Q\bar{Q}$ configuration in non-relativistic quark model. The theoretical results of charmed quark-antiquark spectra, as a sample of heavy mesons, are given in the forth section which are compared with experiments. Additional to that the charmed quark-antiquark bound states wave functions are illustrated in the same section. Eventually, some concluding remarks are presented in the last section.

## 2. The model

Since the discovery of the first quark-antiquark system, namely the charmonium states J/ψ and ψ` in November revolution 1974, the charmonium system has become the prototypical the exotic atom positronium $e^-e^+$ of meson spectroscopy [20-22]. It is found that, the correlation between charmonium states and those of positronium is almost perfect [17], [23, 24]. In the aftermath, many quark-antiquark systems have been discovered. Some of them were produced from $e^-e^+$ annihilation beyond generation of a virtual photon [25-27]. Similarity and correlation make people call quark-antiquark system by Quarkonium. Additional to that, similarity leads to refine a hydrogen or positronium-like potential that can be applied to heavy mesons in non-relativistic QCD as well.

Our scheme in the current research is essentially a potential approach. The mesonic dynamics, in our non-relativistic approximation, is governed by a Hamiltonian which is composed of two parts: a kinetic energy term T and potential energy term V which takes into account the phenomenological interaction between the quark and the antiquark.

$$\hat{H} = \hat{T} + \hat{V} \tag{1}$$

Besides these fundamental ingredients, the mesonic wave functions are illustrated as the eigenfunctions of Schrödinger equation.

$$\hat{H}\psi = E\psi \tag{2}$$

### 2.1. Kinetic energy

The non-relativistic expression of kinetic energy term is given by [9]:

$$\hat{T}_{nr} = m_Q + m_{\bar{Q}} + \frac{\hat{P}}{\mu} \tag{3}$$

where $m_Q$ and $m_{\bar{Q}}$ are respectively the constituent masses of the quark and the antiquark. $\mu$ is the reduced mass for heavy meson system and $\hat{P}$ is the relative momentum. Such a term is expected to give good results for heavy Quarkonium. The former term have been adjusted by the constituent masses to make the calculations of Quarkonium properties could be done and compared to the experiments to see how our model work [8]. Moreover, the center of mass motion is certainly treated in the $\hat{T}_{nr}$ expression.

### 2.2. Potential energy

As stressed in the introduction, our aim is to compare two different forms for $Q\bar{Q}$ potential term. The preliminary attempt to refine positronium-like potential, to get a reasonable static potential for quarkonium, is commonly named a color Coulombic-like potential.

Intuitively, the refined Coulombic-like potential relies on QCD language. It corresponds to the quark interaction at short distance [28, 29]. According to the Quantum Chromodynamics theory, the short-distance behavior is dominated by one-gluon exchange [30, 17]:

$$V_{Coulomic-like}(r) = \frac{-3}{4} \frac{\alpha_s}{r} \tag{4}$$

where $\alpha_s$ is the Chromodynamics analog to hydrogen atom fine-structure, which is so-called the running coupling constant and $\frac{-3}{4}$ is the appropriate color factor.

Phenomenologically, the quark confinement should be accounted at long distances [8-9]. Confinement is an important feature of strong interactions that is widely accepted. According to Chromodynamics, the confining long-range term of $Q\overline{Q}$ potential increases, linearly, with the interquark separation [17], [31]. Consequently, the basic confinement potential is certainly given by:

$$V_{Conf}(r) = br \tag{5}$$

Ultimately, the standard color Coulombic-like plus linear scalar form of $Q\overline{Q}$ potential is simply given by

$$V_1(r) = \frac{-3}{4} \frac{\alpha_s}{r} + br \tag{6}$$

The parameters $\alpha_s$ and $b$ are determined by fitting the corresponding spectrum of each quarkonium states (see the parameters used with different potentials for charmonium in table 1) [28].

The second form of $Q\overline{Q}$ potential model includes a Gaussian-smeared contact hyperfine interaction in the zero-order potential accordingly to study at Oak Ridge national laboratory [32]. Eventually, the second form of quarkonium central potential is:

$$V_2(r) = \frac{-3}{4} \frac{\alpha_s}{r} + br + \frac{32 \pi \alpha_s}{9 m_Q m_{\overline{Q}}} \delta_\sigma(r) \vec{S}_Q . \vec{S}_{\overline{Q}} \tag{7}$$

where $\delta_\sigma(r) = \left(\frac{\sigma}{\sqrt{\pi}}\right)^3 e^{-\sigma^2 r^2}$. $\sigma$, $m_Q$, and $m_{\overline{Q}}$ are also determined by fitting the spectrum. The spin-spin contact hyperfine interaction term $\vec{S}_Q . \vec{S}_{\overline{Q}}$ is one of the spin-dependent terms predicted by OGE (One Gluon Exchange) forces [9], [32], [24].

Actually, spin-dependent interquark forces are evident in the splitting of states within $Q\overline{Q}$ multiplets. They are consistent with the predictions of OGE Breit-Fermi Hamiltonian, combined with the guess-work of Lorentz scalar confining interaction [32, 33].

Now, we want to use the previous two forms as minimal potential models to get the spectra of charmonium ($c\bar{c}$) bound states, with wavefunctions determined by the radial Schrödinger equation.

**Table 1.** The parameters used in different potentials to fit the result masses of $c\bar{c}$ states in $GeV$ according to QCD theory.

| parameters | Theo. (NR) | Theo. (NR) |
|:---:|:---:|:---:|
| | V1 potential | V2 potential |
| $m_Q = m_{\bar{Q}}$ (GeV) | 1.4495 | 1.4399 |
| $\alpha_s$ | 0.5317 | 0.4827 |
| b (GeV)$^2$ | 0.1497 | 0.1488 |
| $\sigma$ (GeV) | —— | 1.2819 |

### 3. Matrix method to solve the radial Schrödinger equation for $Q\bar{Q}$ central potentials

As stressed in the previous section, our approach bases on studying the mass spectra of charmonium $c\bar{c}$ states through solving the Schrödinger equation with the advanced introduced $Q\bar{Q}$ central potentials.

Various symplectic numerical schemes are commonly used to solve the time-independent Schrödinger equation [34]. One of the symplectic matrix schemes is extended to the solution of time-independent Schrödinger equation in spherical symmetric $Q\bar{Q}$ potentials [23, 24]. We refer to this scheme as Matrix method.

Two body charm-anticharm problem has to be reduced to one problem with reduced mass $\mu$. The associated dispersion distance from a specific point in the non-relativistic model is $\vec{r}$. The mass of our reduced particle can be defined as:

$$\mu = \frac{m_c m_{\bar{c}}}{m_c + m_{\bar{c}}}$$

The Hamiltonian in equation 2 is spherically symmetric. Therefore, the radial Schrödinger equation can be used to determine the charmonium $c\bar{c}$ bound states wavefunctions. It is represented as:

$$\frac{-1}{2\mu} \frac{\partial^2}{\partial r^2} U(r) + \left( \frac{1}{2\mu r^2} l(l+1) + [V(r) - E] \right) U(r) = 0 \qquad (8)$$

where $U(r)$ is the radial wavefunctions.

Our perspective relies on solving equation 8 numerically as a matrix eigenvalue problem. For the sake of simplicity, the finite difference approximation of radial second derivative can be transformed into tridiagonal matrix form. The finite difference quotient of the second derivative could be expressed as:

$$\frac{d^2 U(r)}{dr^2} = \frac{U_{i+1} - 2U_i + U_{i+1}}{d^2} + \mathcal{O}(\Delta r^3)$$

where $d$ is the uniform mesh spacing of our computational grid. $i$th points are the sites on a computational grid corresponds to position $r_i$ where $i = 1, 2, 3, \ldots, (N-1)$ and $N$ is the number of steps all over the range of the grid. The computational sampling width is:

$$d = \frac{R_{max} - R_{min}}{N}$$

Where $R_{max}$ and $R_{min}$ are respectively the maximum and minimum values of the radial distance. Thus, we can define any arbitrary value of $r_i$ as:

$$r_i = R_{min} + id$$

Eventually, the approximation can be done by rewrite the radial Schrödinger equation for discrete space $r_i$ as:

$$C_2 U_{i+1} + \left(C_1 + V(r_i) + \frac{1}{2\mu}\frac{l(l+1)}{r_i}\right)U_i + C_2 U_{i-1} = EU_i \qquad (9)$$

where $C_1 = \frac{1}{\mu d^2}$ and $C_2 = \frac{-1}{2\mu d^2}$ . Additionally, equation 9 can be reduced by considering:

$$h(i) = C_1 + V(r_i) + \frac{1}{2\mu}\frac{l(l+1)}{r_i^2}$$

Ultimately, the matrix method algorithm is obtained as:

$$C_2 U_{i+1} + h(i)U_i + C_2 U_{i-1} = EU_i \qquad (10)$$

Consequently, the general symplectic form of the matrix method can be representing as a tridiagonal matrix of dimension $(N-1) \times (N-1)$:

$$\begin{pmatrix} h(1) & C_2 & 0 & 0 & ... & 0 \\ C_2 & h(2) & C_2 & 0 & ... & ... \\ ... & ... & ... & ... & ... & ... \\ 0 & ... & ... & ... & h(N-2) & C_2 \\ 0 & ... & ... & ... & C_2 & h(N-1) \end{pmatrix} \begin{pmatrix} U_1 \\ U_2 \\ ... \\ ... \\ U_{N-1} \end{pmatrix} = E \begin{pmatrix} U_1 \\ U_2 \\ ... \\ ... \\ U_{N-1} \end{pmatrix} \qquad (11)$$

Through the previous scheme, we can solve the eigenvalue problem of any arbitrary quark-antiquark bound states.

## 4. Results and discussion

Two proposed phenomenological potential models are used to study the mass spectra of charmonium bound states, numerically, through using matrix method. Wavefunctions and the corresponding eigenvalues are calculated by the same method.

It is crucial to use a proper numerical scheme, for calculating the meson mass spectra, with highest possible accuracy and the least computer time. From this standpoint, it is important to compare our results, corresponding to the spectra, with experimental data. All charmonium resonances chosen in our comparison are taken from the review of Particle Data Group (PDG) experiments [27].

In our approach, the accuracy of both matrix scheme and the used potential models is determined numerically by minimizing a $\chi^2$ function defined by

$$\chi^2 = \frac{1}{n} \sum_{i=1}^{n} \frac{(M_i^{exp} - M_i^{th})}{\Delta_i} \tag{12}$$

In the previous formula, the summation runs over selected samples n of $c\bar{c}$ meson states. $M_i^{exp}$ is the experimental mass of charmonium, while $M_i^{th}$ is the corresponding theoretical mass depending upon the potential used. The $\Delta_i$ quantity is the experimental uncertainty on the mass. Intuitively, $\Delta_i$ should be one. The tendency of overestimating $\chi^2$ value is that, it reflects some mean error per heavy meson state.

The normalized radial wave functions of charmonium states are illustrated for each used potential in figures [1-6]. Predicted $c\bar{c}$ masses and a value $\chi^2$ obtained are represented in table 2. We observe that the various used potentials can give appreciable $\chi^2$ values by using matrix method for various grid spaces.

In the first potential, we calculated the $\chi^2$ value, using equation 12, and we obtained small results that suggest the proper choice of the potential. For the second potential, the computed $\chi^2$ value is improved in such a way that favors the choice of second potential over the first potential.

In order to study the optimal grid size, which results in correct $\chi^2$ values, we attempted four different grids in our calculations per potential (see table 2).

We found that with the matrix method, we can go as low as $(50 \times 50)$ in the grid size without losing a substantial accuracy. It is gratifying to find that, there are slight differences in $\chi^2$ values for all used computational grid sizes. A stable $\chi^2$ value occurred at grid size $(50 \times 50)$ and continued for the higher gird spaces. Obviously, the computed $\chi^2$ value is small, for all cases, to the limit that we trust our approach. The first potential model, i.e., $V_1$ does not distinguish between the states that have a similar angular momentum. Meanwhile, the second potential $V_2$ does distinguish between the states that have similar angular momentum as it uses the intrinsic angular momentum.

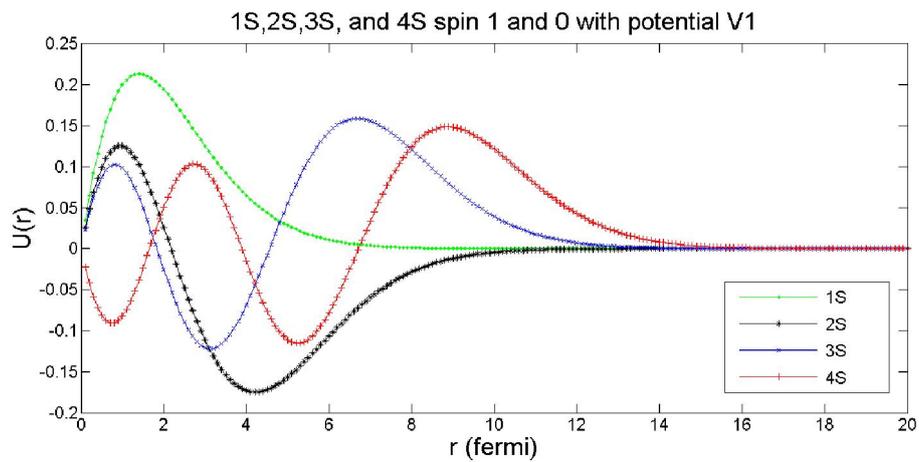

**Figure 1: The wavefunctions of S states for potential one.**

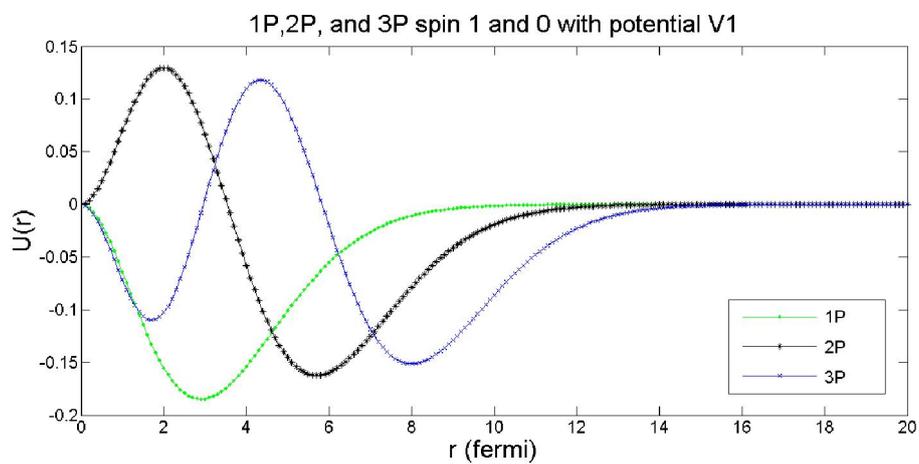

**Figure 2: The wavefunctions of P states for potential one**

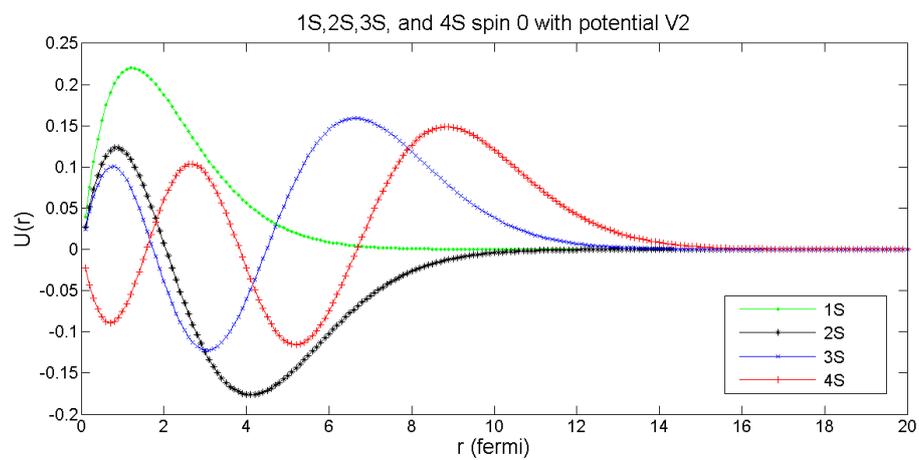

**Figure 3: The wavefunctions of S states spin zero for potential two**

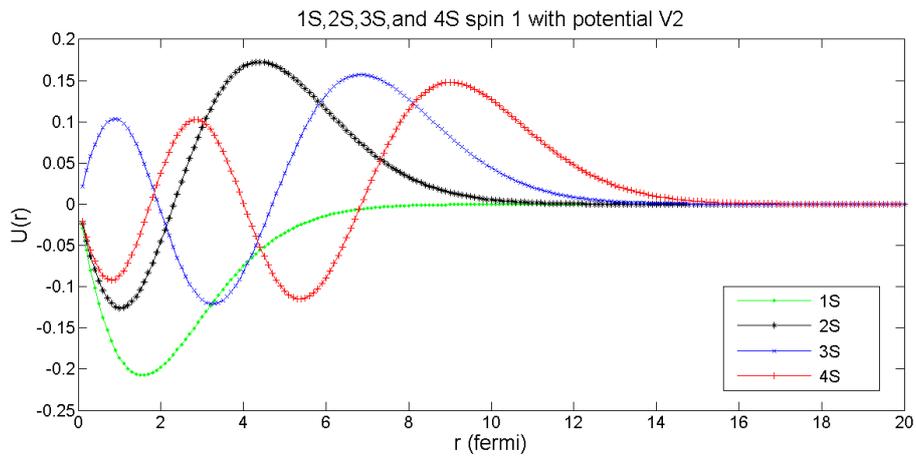

**Figure 4: The wavefunctions of S states spin one for potential two**

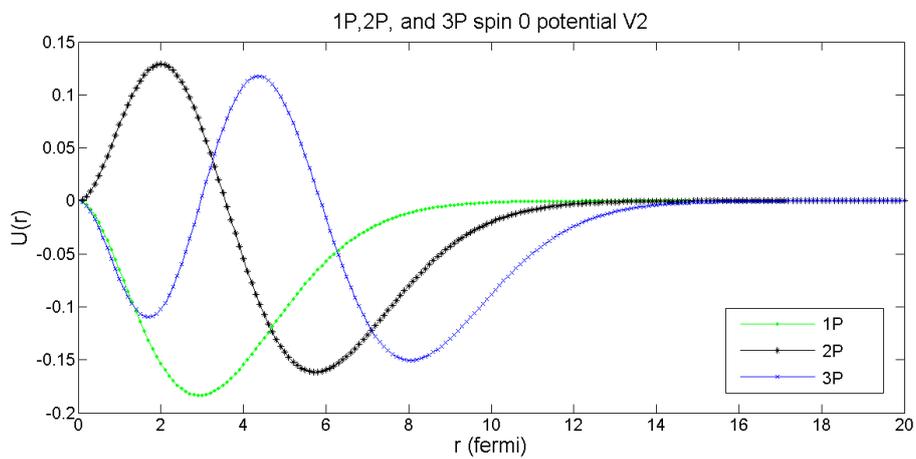

**Figure 5: The wave-functions of P states spin zero for potential two**

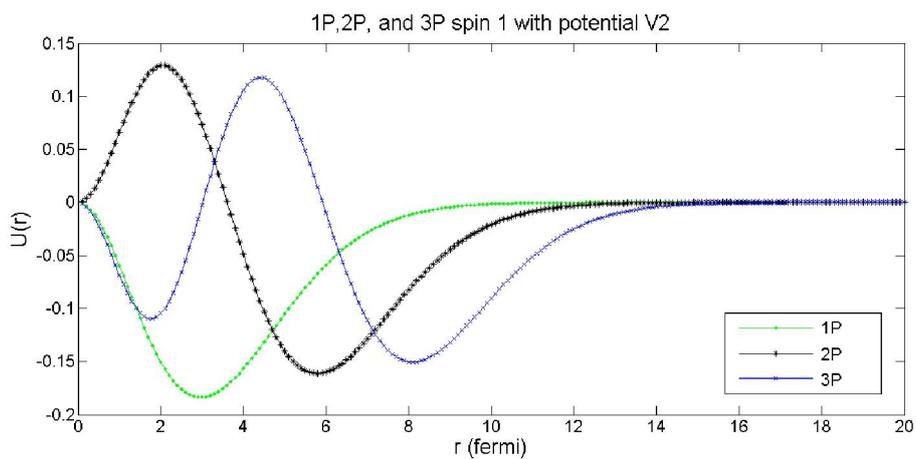

**Figure 6: The wave-functions of P states spin one for potential two**

**Table 2.** Experimental and theoretical spectra of [$c\bar{c}$] states, in GeV, for various computational grid number. The considered $c\bar{c}$ radius is equal 20 Fermi.

| state | name | Exp. Mass [27] | Theoretical masses for grid $(30 \times 30)$ | | Theoretical masses for grid $(50 \times 50)$ | | Theoretical masses for grid $(100 \times 100)$ | |
|---|---|---|---|---|---|---|---|---|
| | | | $V_1$ | $V_2$ | $V_1$ | $V_1$ | $V_1$ | $V_2$ |
| $1^3S_1$ | $J/\psi$ | $3.9687 \pm 0.0004$ | 3.0496 | 3.0998 | 3.0448 | 3.0970 | 3.0429 | 3.0958 |
| $1^1S_0$ | $\eta_c(1S)$ | $2.9792 \pm 0.0013$ | 3.0496 | 2.9924 | 3.0448 | 2.9885 | 3.0429 | 2.9882 |
| $2^3S_1$ | $\psi(2S)$ | $3.68609 \pm 0.0004$ | 3.6366 | 3.6552 | 3.6433 | 3.6618 | 3.6462 | 3.6645 |
| $2^1S_0$ | $\dot{\eta}(2S)$ | $3.637 \pm 0.0004$ | 3.6366 | 3.6033 | 3.6433 | 3.6118 | 3.6462 | 3.6160 |
| $3^3S_1$ | $\psi(3S)$ | $4.039 \pm 0.001$ | 4.0391 | 4.0482 | 4.0540 | 4.0625 | 4.0603 | 4.0684 |
| $3^1S_0$ | $\eta_c(3S)$ | | 4.0391 | 4.0098 | 4.0540 | 4.0260 | 4.0603 | 4.0334 |
| $4^3S_1$ | $\psi(4S)$ | $4.421 \pm 0.0004$ | 4.3737 | 4.3777 | 4.3971 | 4.4002 | 4.4066 | 4.4093 |
| $4^1S_0$ | $\eta_c(4S)$ | | 4.3737 | 4.3460 | 4.3971 | 4.3703 | 4.4066 | 4.3809 |
| $1^3P_2$ | $\chi_2(1P)$ | $3.55620 \pm 0.00009$ | 3.4947 | 3.5003 | 3.4984 | 3.5037 | 3.4999 | 3.5051 |
| $1^3P_1$ | $\chi_2(1P)$ | $3.51066 \pm 0.00007$ | 3.4947 | 3.5003 | 3.4984 | 3.5037 | 3.4999 | 3.5051 |
| $1^3P_0$ | $\chi_0(1P)$ | $3.41475 \pm 0.00031$ | 3.4947 | 3.5003 | 3.4984 | 3.5037 | 3.4999 | 3.5051 |
| $1^1P_1$ | $h_c(1P)$ | $3.52541 \pm 0.00016$ | 3.4947 | 3.4932 | 3.4984 | 3.4970 | 3.4999 | 3.4985 |
| $2^3P_2$ | $\chi_2(2P)$ | | 3.9188 | 3.9184 | 3.9274 | 3.9265 | 3.9310 | 3.9298 |
| $2^3P_1$ | $\chi_2(2P)$ | | 3.9188 | 3.9184 | 3.9274 | 3.9265 | 3.9310 | 3.9298 |
| $2^3P_0$ | $\chi_0(2P)$ | | 3.9188 | 3.9184 | 3.9274 | 3.9265 | 3.9310 | 3.9298 |
| $2^1P_1$ | $h_c(2P)$ | | 3.9188 | 3.9098 | 3.9274 | 3.9186 | 3.9310 | 3.9221 |
| $3^3P_2$ | $\chi_2(3P)$ | | 4.2670 | 4.2633 | 4.2818 | 4.2775 | 4.2879 | 4.2833 |
| $3^3P_1$ | $\chi_2(3P)$ | | 4.2670 | 4.2633 | 4.2818 | 4.2775 | 4.2879 | 4.2833 |
| $3^3P_0$ | $\chi_0(3P)$ | | 4.2670 | 4.2633 | 4.2818 | 4.2775 | 4.2879 | 4.2833 |
| $3^1P_1$ | $h_c(3P)$ | | 4.2670 | 4.2541 | 4.2818 | 4.2691 | 4.2879 | 4.2751 |
| $1^3D_3$ | $\psi_3(1D)$ | | 3.7824 | 3.7787 | 3.7839 | 3.7801 | 3.7845 | 3.7808 |
| $1^3D_2$ | $\psi_2(1D)$ | | 3.7824 | 3.7787 | 3.7839 | 3.7801 | 3.7845 | 3.7808 |
| $1^3D_1$ | $\psi(1D)$ | $3.77292 \pm 0.00035$ | 3.7824 | 3.7787 | 3.7839 | 3.7801 | 3.7845 | 3.7808 |
| $1^1D_2$ | $\psi_{c2}(1D)$ | | 3.7824 | 3.7783 | 3.7839 | 3.7797 | 3.7845 | 3.7804 |
| $2^3D_3$ | $\psi_3(2D)$ | | 4.1482 | 4.1412 | 4.1533 | 4.1462 | 4.1554 | 4.1482 |
| $2^3D_2$ | $\psi_2(2D)$ | | 4.1482 | 4.1412 | 4.1533 | 4.1462 | 4.1554 | 4.1482 |
| $2^3D_1$ | $\psi(2D)$ | $4.153 \pm 0.0003$ | 4.1482 | 4.1412 | 4.1533 | 4.1462 | 4.1554 | 4.1482 |
| $2^1D_2$ | $\psi_{c2}(2D)$ | | 4.1482 | 4.1405 | 4.1533 | 4.1455 | 4.1554 | 4.1475 |
| | $\chi^2$ | | 0.0019 | 0.0013 | 0.0017 | 0.0011 | 0.0017 | 0.0011 |

## 5. Conclusions

In our approach, we proposed two different models of Hamiltonian to describe the spectra of heavy mesons. The first potential includes two phenomenological contributions, which are the coulomb one-gluon exchange and the linear confinement. On the other hand, the second model is extended by the hyperfine term. Our expression of the hyperfine term is based phenomenologically on spin-spin interactions of quarkonium constituents.

In this paper, our approach compares to the methodologies developed by two distinct groups [7-8]. However, it differs in numerous points. In particular, they tried to stick more to the basis of QCD. Moreover, they added tensor and spin-orbit terms to split a number of multiplets which seems to us very complicated. Actually, spin-orbit and tensor terms will be included in the next step by using our computational strategy. It will be an attempt towards more sophisticated treatments of the meson sectors. Such attempt is in progress. Additionally, the previous groups used a very time consuming task to describe mesons. Eventually, they did not restrict them self on a non-relativistic frame work leading to increase the complexity.

Our philosophy relies on proposing good potential models which correctly describe the overall properties of quark-antiquark pairs and whose form is simple enough to be used in much more complicated systems. We restrict ourselves on studying the spectra and the associated wave function as a preliminary step to more sophisticated steps in the near future following this approach. Our potential models are already a very good approximation for handling the meson sector. Both of them give a satisfactory agreement with experiments. Desired $\chi^2$ values are occurred for each of them.

We observed that, the first potential $V_1$ is a blind model as it cannot distinguish among the multiplets that have a similar angular momentum. In contrast, the second potential $V_2$ does distinguish between them because it uses the intrinsic angular momentum $\vec{S}$.

Our numerical strategy is based on using the finite difference algorithm in matrix form for making the calculations of a full spectra and the wave function associated very fast and very easy.
Studying the full spectra of $Q\bar{Q}$ through matrix method allow to use a computational grid as low as $(50 \times 50)$ without losing the substantial minimization corresponding to the $\chi^2$ value. Ultimately, the description of heavy mesons is better with using both V2 and matrix method.

### Acknowledgments

This work has been partially funded by AIMS-NEI studentship which is gratefully acknowledged. M. S. Ali would like to acknowledge AIMS-Senegal committee for the continued support.